\documentclass[a4paper,fleqn,12pt]{cas-sc}

\usepackage{textcomp}
\usepackage[numbers]{natbib}
\usepackage{amsmath}
\usepackage{chemformula}

\newcommand{\highlight}[1]{{#1}}

\ExplSyntaxOn
\cs_set:Npn \__first_footerline: {}
\ExplSyntaxOff

\begin{document}

\let\WriteBookmarks\relax
\def\floatpagepagefraction{1}
\def\textpagefraction{.001}

\shorttitle{Spectroscopy of the C$_{\rm 3v}$(O$^{2-}$) Centre in CaF$_2$:Er$^{3+}$}    

\shortauthors{M.D. Moull et. al.}  

\title [mode = title]{Spectroscopy, Crystal-Field, and Transition Intensity Analyses of the C$_{\rm 3v}$(O$^{2-}$) Centre in Er$^{3+}$ Doped CaF$_{2}$ Crystals}  

\author[1,2]{M. D. Moull}[]
\credit{Investigation, Formal analysis, Visualization, Writing – original draft}
\author[1,2]{J. B. L. Martin}[]
\credit{Investigation, Formal analysis, Supervision}
\author[3,4]{T. G. M. Newman}[]
\credit{Investigation, Writing - review and editing}
\author[3]{A. L. Jeffery}[]
\credit{Investigation, Writing - review and editing}
\author[3,4]{J. G. Bartholomew}[]
\credit{Conceptualization, Investigation, Supervision, Writing - review and editing}
\author[1,2]{J.-P. R. Wells}[orcid=0000-0002-8421-6604]
\cormark[1]
\credit{Supervision, Writing - review and editing, Visualization, Resources, Funding Acquisition, Project Administration}
\ead{jon-paul.wells@canterbury.ac.nz}
\author[1,2]{M. F. Reid}[orcid=0000-0002-2984-9951]
\cormark[1]
\credit{Conceptualization, Software, Investigation, Supervision, Writing - review and editing}
\ead{mike.reid@canterbury.ac.nz}

\affiliation[1]{organization={The School of Physical and Chemical Sciences, University of Canterbury},
            addressline={PB4800 Christchurch 8140, New Zealand}
           }

\affiliation[2]{organization={The Dodd-Walls Centre for Photonic and Quantum Technologies},
            addressline={New Zealand}
           }

\affiliation[3]{organization={Centre for Engineered Quantum Systems},
            addressline={School of Physics, The University of Sydney}, 
         citysep={}, 
           state={NSW},
            country={Australia}
           }

\affiliation[4]{organization={The University of Sydney Nano Institute},
            addressline={The University of Sydney}, 
         citysep={}, 
           state={NSW},
            country={Australia}
           }

\cortext[1]{Corresponding authors}

\fntext[1]{}

\begin{abstract}
  Erbium ions in crystals show considerable promise for the technologies that will form the backbone of future networked quantum information technology. Despite advances in leveraging erbium's fibre-compatible infrared transition for classical and quantum applications, the transitions are, in general, not well understood. We present detailed absorption and laser site-selective spectroscopy of the C$_{\rm 3v}$(O$^{2-}$) centre in CaF$_2$:Er$^{3+}$ as an interesting erbium site case study. The $^{4}$I$_{15/2}$Z$_1 \rightarrow {^{4}}$I$_{13/2}$Y$_1$ transition has a low-temperature inhomogeneous linewidth of 1 GHz with hyperfine structure observable from the  $^{167}$Er isotope. A 
\highlight{
parametrized crystal-field Hamiltonian
}
is fitted to 34 energy levels and the two ground state magnetic splitting factors. The wavefunctions are used to perform a transition intensity analysis and electric-dipole parameters are fitted to absorption oscillator strengths. Simulated spectra for the $^{4}$I$_{11/2}\rightarrow {^{4}}$I$_{15/2}$ and $^{4}$I$_{13/2} \rightarrow {^{4}}$I$_{15/2}$ inter-multiplet transitions are in excellent agreement with the experimentally measured spectra.  The  $^{4}$I$_{13/2}$ excited state lifetime is 25.0\,ms and the intensity calculation is in excellent agreement with this value.
\\

\end{abstract}





\maketitle

\section{Introduction}\label{intro}

Rare-earth doped crystals are a strong candidate material for quantum information technologies that interface between light and matter, such as quantum memories, repeaters, and transducers~\cite{thiel2011rare, goldner2015rare, Becher2023}. 
In addition to high performance ensemble-based technologies, single rare-earth ions also show promise as spin-photon interfaces for entanglement distribution~\cite{ruskuc2024, uysal2024}.
  Trivalent erbium (Er$^{3+}$) is of particular interest due to its compatibility with existing optical fiber technologies as the frequencies of transitions between the lowest lying $^{4}$I$_{15/2}$ and $^{4}$I$_{13/2}$ multiplets are in, or near, the infrared telecom-C band~\cite{raha2021, stevenson2022erbium}.
An ongoing challenge is that the $^{4}$I$_{15/2} \leftrightarrow ^{4}$I$_{13/2}$ transitions of interest, like most 4f-4f optical transitions of rare-earths in solid state hosts, are weak.
The weakly allowed electric-dipole transitions are of similar strength to the magnetic-dipole transitions, with dipole moments approximately 1000 times less than other solid state emitters such as colour centres and quantum dots. 
Small optical dipole moments limit emission rates and transition cyclicity, and impede strongly coupling these transitions to optical cavities, all of which restrict the performance of single rare-earth ions for quantum computing and communication applications. 
The magnetic- and electric-dipole moments are influenced by the crystal-field environment of the ion in the crystalline host at a specific site. Thus, understanding these crystal-field perturbations is important to understand the fundamental limits on erbium infrared transition strengths. 


Crystal-field modelling is currently used to calculate rare-earth ion electronic structure based on existing spectroscopic data. 
These phenomenological models have been extremely successful in performing parametrized crystal-field calculations for a variety of rare-earth ions / host combinations, including popular host yttrium orthosilicate (\ch{Y2SiO5}), despite the significant challenges posed by the substitutional sites' low point-group symmetry~\cite{sebastian, transfer, zhou_2020, yasharce, yasharnd, nicksm, nicker}.
Single-erbium control has been achieved in this material, in part motivated by the suggestion that optical  dipole moments at low symmetry \ch{Y2SiO5} sites would be among the highest possible for rare-earth hosts~\cite{McAuslan2009, dibos2018}.
However, it has been shown that other sites in other hosts may have higher dipole moments~\cite{thiel2011rare, Xie2021}, and moreover that there may be other hosts that are more suitable for single-ion technologies when optimization of the overall ion-host-cavity system is considered~\cite{stevenson2022erbium, ourari2023}.



Given the opportunities for further optimising the combination of host material and ion site to achieve the best properties for a rare-earth based spin-photon interface, it would be desirable to perform \textit{ab initio} crystal-field calculations to predict the best materials to host rare-earths for single-ion technologies. 
However, such calculations have so far been prohibitively challenging, especially in systems where the rare-earth ion occupies low point-group symmetry sites, as in \ch{Y2SiO5}. 
Much of the field’s understanding of how the crystal field in a specific host and site will affect an ion’s properties relies on rules of thumb which have not been systematically substantiated.
By closely studying the electronic structure of ions in sites where the crystal-field is well-understood - such as those with higher symmetry - a better understanding could be developed of precisely how the local environment affects an ion’s properties. 
Eventually, a combination of crystal-field modelling and advanced fabrication techniques could be used to predict and then create sites for rare-earths with optical properties that have been optimized for quantum technology applications. 

\section{Motivation to study calcium fluoride}\label{caf2}
The alkaline earth fluorides afford a variety of high symmetry crystallographic sites that vary from host to host and depend upon any post growth sample treatment. 
In these materials, it has already been demonstrated that complex Zeeman-hyperfine structure, including intensities, can be modelled \cite{wells, hozeeman}. 

Calcium fluoride has a cubic structure with space group Fm$\overline{3}$m. The cubic cage is formed by eight F$^{-}$ ions, with the Ca$^{2+}$ ion occupying the centre of each alternate cage. 
Trivalent erbium ions readily substitute into the lattice during growth. When they replace a Ca$^{2+}$ cation,  some form of charge compensation is required to maintain overall charge neutrality. 
For crystals grown in an inert environment or under vacuum, this takes the form of an interstitial F$^{-}$ ion, yielding centres of tetragonal C$_{\rm 4v}$(F$^{-}$), trigonal C$_{\rm 3v}$(F$^{-}$) symmetry, as well as a non-locally charge compensated cubic (O$_{\rm h}$) symmetry centres \cite{tallant1975selective}. 

\highlight{ 
Charge compensation may also be achieved by the substitution of oxygen (O$^{2-}$) ions for 
F$^{-}$ ions. 
In precipitates, several oxygen-compensated sites have been reported, including one where four O$^{2-}$ ions substitute for seven F$^{-}$ ions \cite{gustafson1979trace}. 
However, in bulk single crystals the predominant O$^{2-}$ centre arises from the  substitution of a single nearest-neighbour F$^{-}$ ion by an O$^{2-}$ ion.  
This centre has C$_{\rm 3v}$ point-group symmetry \cite{ranon1963electron, Cockroft90, chehaidar2000interstitial}.
}
Previous crystal-field analyses of this centre (which we denote as the C$_{\rm 3v}$(O$^{2-}$) centre) in both Er$^{3+}$ and Eu$^{3+}$ doped CaF$_{2}$ show very large axial crystal-field parameters \cite{Cockroft90, chehaidar2000interstitial, smith2022complete} and, in the case of Eu$^{3+}$, significantly increased transition intensities \cite{silversmith1985zeeman, gustafson1979trace} with an anomalously high dipole moment \cite{Bartholomew}. 
Understanding how the Er$^{3+}$ dipole moment is impacted by this strongly axial crystal field in the Er$^{3+}$ C$_{\rm 3v}$(O$^{2-}$) site will yield additional insights relevant for single ion manipulation in the 1.55~$\mu$m region. Moreover, comparing the Er$^{3+}$ to the Eu$^{3+}$ site will generate useful insight towards our ultimate goal of more confidently predicting and engineering the optical properties of rare-earths in under-explored host systems.

 
\section{Theory}\label{theory}

\subsection{The Crystal-Field Hamiltonian}
The 4f$^{11}$ configuration of Er$^{3+}$ in a host crystal can be modelled by a parametrized
\highlight{
crystal-field
}
Hamiltonian \cite{liu2005}:
\begin{equation}
    H = H_\text{FI} + H_\text{CF},
\end{equation}
where $H_\text{FI}$ is the free-ion contribution and $H_\text{CF}$ is the crystal-field contribution. The crystal-field contribution can be expanded in terms of spherical tensors as:
\begin{equation}
\label{eq:cf}
    H_\text{CF} = \sum_{k,q} B_{q}^{k}C_{q}^{(k)}.
\end{equation}
The  $B_{q}^{k}$ are crystal-field parameters and the $C_{q}^{(k)}$ are  Racah spherical tensor operators acting within the 4f$^N$ configuration. 

The Er$^{3+}$ oxygen centre discussed in this work has C$_{\rm 3v}$ point-group symmetry so only parameters with $q =0$, 3, and 6 are non-zero. The crystal-field Hamiltonian may therefore be written as 
\begin{eqnarray}\label{eq:c3v_cf}
  H_{CF} &=& B_{0}^{2}C_{0}^{2}+B_{0}^{4}C_{0}^{4}+B_{3}^{4}(C_{-3}^{4}-C_{3}^{4}) \nonumber \\
     &+&B_{0}^{6}C_{0}^{6}
     +B_{3}^{6}(C_{-3}^{6}-C_{3}^{6})
     +B_{6}^{6}(C_{-6}^{6}+C_{6}^{6}).
\end{eqnarray}

\subsection{Transition Intensities}

Transition intensities for lanthanide ions are dominated by electric-dipole (ED) and magnetic-dipole (MD) interactions. Details of notation are given in  \cite{Reid2006spectroscopic,ReRi83c,ReDaRi83}.
\highlight{
Magnetic-dipole moments may be accurately calculated for transitions within the 4f$^N$ configuration by evaluating the operator 
\begin{equation}
    M_{q}^{(1)}=\frac{-e\hbar}{3mc}(L_{q}^{(1)}+S_{q}^{(1)})
  \end{equation}
  between the crystal-field eigenstates.
  Here $q$ represents the polarization, which can have values of $q=0,\pm 1$ (or linear  combinations for appropriate polarization directions). 
On the other hand, the electric-dipole operator,  
\begin{equation}
    -eD_{q}^{(1)}=-erC_{q}^{(1)}  ,
  \end{equation}
  vanishes for transitions within the 4f$^N$ configuration.  Electric-dipole transitions are only allowed due to mixing with opposite-parity configurations on the ion (``static coupling'') or coupling to dipoles induced on the ligands by the radiation field (``dynamic coupling'' or ``ligand polarization''). These mechanisms will be discussed further in Section \ref{sec:predictions}. 
  To parametrize these various effects, we  define an \emph{effective} dipole moment
}
\begin{equation}
\label{eq:inten} D_{\text{eff},q}=\sum_{\lambda,t,p}A_{tp}^{\lambda}U_{p+q}^{\lambda}(-1)^{q}\langle\lambda~(p+q),1~{-q}\vert tp\rangle,
\end{equation}
where $\lambda=2,4,6$, $t = \lambda-1,\lambda,\lambda+1$ and $p$ is dependent on the point-group symmetry of the crystal-field, $A_{tp}^{\lambda}$ are parameters for the particular site, and $U_{p+q}^{\lambda}$ are unit tensors \cite{Reid&Richardson}.

We define  electric- and magnetic-dipole line strengths for a transition between
initial and final states $I$ and $F$ with polarization $q$: 
\begin{equation} \label{eq:edstrength}
S_{FI,q}^{\rm ED}=\sum_i\sum_f e^2\left| \langle Ff |
D_q^{(1)} | Ii\rangle \right| ^2  , 
\end{equation}
\begin{equation} \label{eq:mdstrength}
S_{FI,q}^{\rm MD}=\sum_i\sum_f \left| \langle Ff| M_q^{(1)} |
Ii\rangle \right| ^2  .
\end{equation}
The sums in these equations are over all components of the initial and
final states. 

For absorption measurements, observable quantities are related to oscillator strengths. The oscillator strength may be calculated from the dipole strengths by
\begin{eqnarray}\label{eq:f}
f_{FI,q} &=& 
f^{\rm ED}_{FI,q} + f^{\rm MD}_{FI,q}     \nonumber \\
&=& \frac{2 m \omega_{FI}}{\hbar e^2}\frac{1}{g_I}
\left(
                         \frac{\chi_{\rm L}}{n} 
                         S^{\rm ED}_{FI,q}
+
                         {n} 
                         S^{\rm MD}_{FI,q}
\right).
\end{eqnarray}
These expressions include the degeneracy of the initial state, $g_I$, the refractive index, $n$, and the local electric-field correction. Here we assume the Lorentz or virtual cavity model correction given by 
\begin{equation}\label{eq:chi}
\chi_{\rm L} = \left( \frac{n^2+2}{3} \right)^2. 
\end{equation}

For unpolarized measurements, the oscillator strengths for different polarizations must be averaged, to give 
\begin{equation}\label{eq:faverage}
\bar{f}_{FI} = \frac{1}{3}\sum_q \left(  f_{FI,q}  \right).
\end{equation}
For spontaneous emission, the radiative lifetime of state $I$,  $\tau_{I, \text{radiative}}$, is the inverse of the sum of the Einstein $A$ coefficients for decay to all states of lower energy: 
\cite{Reid2006spectroscopic}: 
\begin{eqnarray}\label{eq:radlifetime}
  \frac{1}{\tau_{I,\, \text{radiative}}}
  &=& \sum_q \left( A^{\rm ED}_{FI,q} +  A^{\rm MD}_{FI,q}\right) \nonumber\\
  &=& 
    \sum_{q,F} \frac{1}{3}
  \frac{1}{4 \pi \epsilon_0} \frac{1}{g_I}
    \frac{4 \omega_{FI}^3}{\hbar c^3}
      \left(
                    n \chi_{\rm L} 
                    S^{\rm ED}_{FI,q}
     +  
                     n^3 
                     \frac{1}{g_I}
                     S^{\rm MD}_{FI,q}
                     \right). 
                   \end{eqnarray}

\subsection{Parameter predictions} \label{sec:predictions}

\begin{table}[!tb]
  \begin{center}  
    \caption{Calculated crystal-field and intensity parameters for the C$_{\rm 3v}$(O$^{2-}$) centre.
      Geometry used:
      one O$^{2-}$ at ($r$, $\theta$, $\phi$) $=$ (2.5\,\AA, $0^\circ$, $0^\circ$),
      one F$^-$ at ($r$, $\theta$, $\phi$) $=$ (2.5\,\AA, $70.53^\circ$, $60^\circ$) + C$_3$ operations,
      one F$^-$ at ($r$, $\theta$, $\phi$) $=$ (2.5\,\AA, $109.47^\circ$, $0^\circ$) + C$_3$ operations,
      one F$^-$ at ($r$, $\theta$, $\phi$) $=$ (2.5\,\AA, $180^\circ$, $0^\circ$).
      For O$^{2-}$, $q=-2$, $\alpha = 3.2$\,\AA$^3$.
      For F$^{-}$, $q=-1$, $\alpha = 1.0$\,\AA$^3$.
      Radial integrals used in the calculations were for Eu$^{3+}$ \cite{ReDaRi83}.
    \label{tab:calcparams}}
\bigskip
  
{(a) Calculated crystal-field  parameters (in cm$^{-1}$)
\highlight{
and experimental parameters}.
}
\smallskip

  \begin{tabular}{lll}
  \toprule
    Parameter   &  Point charge & \highlight{Experiment \cite{Cockroft90}}    \\
        \midrule
        $B_{0}^{2}$ &  1735          & 1285\\
        $B_{0}^{4}$ &   489          & 1893\\
        $B_{3}^{4}$ &   394          & 1031\\
        $B_{0}^{6}$ &   130          & 711\\
        $B_{3}^{6}$ &   -60          & -550\\
        $B_{6}^{6}$ &    62          & 632\\
  \bottomrule
  \end{tabular}
  \bigskip

{(b) Calculated electric-dipole transition intensity parameters (in $10^{-12}$\,cm).}
\smallskip

\begin{tabular}{lllll}
\toprule
  Parameter  &\highlight{Static coupling} &\highlight{Dynamic coupling} &Total   \\
        \midrule
        A$^2_{10}$  &   115.1    &      0 &  115.1  \\
        A$^2_{30}$  &    -9.8    &   69.6 &   59.8  \\
        A$^4_{30}$  &   -18.1    &      0 &  -18.1  \\
        A$^4_{50}$  &     3.3    &   -9.1 &   -5.9  \\
        A$^6_{50}$  &    10.0    &      0 &   10.0  \\
        A$^6_{70}$  &    -1.3    &    2.9 &    1.6  \\
  \bottomrule
\end{tabular}
\end{center}
\end{table}

Some insight into the relationship between the local geometry on the crystal-field and intensity parameters may be obtained from a simple point-charge and point-dipole calculation. It is also possible to use the superposition model \cite{NN89a,NeNg00}, but the mixture of ligands in the  C$_{\rm 3v}$(O$^{2-}$) centres makes that complicated in this case. In the following, we 
use the approach of \cite{ReRi83c,ReDaRi83}.

In Table \ref{tab:calcparams} we present calculations using an undistorted cube, but with one of the F$^-$ ligands replaced by an O$^{2-}$. The cube is oriented with the O$^{2-}$ on the $z$ axis.
\highlight{
The coordinates, radial integrals, and polarizabilities used in the calculations are indicated in the Table. 

The crystal-field calculation (Table \ref{tab:calcparams}(a)) treats the ligands as point charges
(\cite{liu2005} Eq.\ 1.56).
We also list the experimental parameters from Ref.\ \cite{Cockroft90}, converted to our notation.
The signs predicted by the point-charge calculation agree with the experimental parameters. However, the magnitudes of the the $k=2$ parameter is overestimated and the $k=6$ parameters are underestimated. This is a well-known shortcoming of point-charge calculations. 
}

\highlight{
  Table \ref{tab:calcparams}(b) gives two contributions to intensity parameters. Ligand charges are used  to estimate the ``static coupling'' contribution from mixing of the 4f$^N$ configuration with other configurations of opposite parity, such as 4f$^N$5d
(\cite{ReDaRi83} Eq.\ 4). Ligand polarizabilities are used to estimate the ``dynamic coupling'' or ``ligand polarization'' contribution, which takes into account dynamic polarization of the ligands by the radiation field (\cite{ReDaRi83} Eq.\ 5). These calculations are not expected to give accurate magnitudes for the parameters. However this approach has had some success in predicting the signs of the parameters. \cite{Reid2006spectroscopic,ReDaRi83}. More sophisticated \emph{ab-initio} calculations have been attempted for a few hosts \cite{ReNgNe89,Wen2014} but accurate \emph{ab-initio} calculation of 4f-4f intensities remains a challenging problem.
}
  
For the undistorted geometry used in this calculation, only intensity parameters with $q=0$ are non-zero. A distortion of the fluoride cage would give parameters with $q=3$ and $q=6$. When comparing fits, it is important to note that a rotation about $z$ by $60^\circ$ would change the sign of the $q=3$ crystal-field parameters and that the intensity parameters can be multiplied by an overall sign without affecting the calculated intensities. 


\section{Experimental Techniques}\label{}

CaF$_2$:Er$^{3+}$ bulk crystals were grown from crushed crystal offcuts obtained from Crystran Ltd. (Poole, United Kingdom) with 0.01 molar \% of ErF$_{3}$ added as the dopant source. The charge was placed inside a graphite crucible and loaded into the work coils of a 38 kW Arthur D.\ Little radio frequency (RF) furnace. The chamber was evacuated to below 10$^{-5}$ Torr and the crucible heated to slightly above 1450 $^{\circ}$C. The crucibles were lowered through the temperature gradient afforded by the single zone furnace over a period of 18 hours. The as-grown crystal boule was cut using a diamond saw into samples approximately 3 mm in width. To introduce oxygen into the crystal lattice, the samples were placed inside a Sentro Tech ST-1800 muffle furnace and heated in an ambient environment at 850 $^{\circ}$C over periods between 5 and 24 hours. An additional sample was also prepared that had been annealed at the higher temperature of 1100 $^{\circ}$C for 5 hours. The heating and cooling rates of the furnace were restricted to
\highlight{
1\,$^{\circ}$C/min.
}
Following annealing, samples were polished with a diamond solution graded to 3 {\textmu}m.

Absorption spectroscopy was performed using a Bruker Vertex 80 Fourier Transform Infrared Spectrometer (FTIR) at a resolution of 0.075 cm$^{-1}$, using samples cooled to a base temperature of 8~K using a JANIS closed-cycle cryostat. For the entirety of the measurements, the optical beam path was purged with N$_{2}$ gas. Fluorescence and excitation spectroscopy was performed using a
\highlight{
Photon Technology International (PTI)
}
tunable dye laser optically pumped by a PTI pulsed nitrogen laser at a 5 Hz repetition rate. Fluorescence spectra were recorded using an iHR550 (Horiba Scientific) spectrometer coupled to a Hamamatsu R2257P thermoelectrically-cooled visible photomultiplier (PMT) tube or a Hamamatsu H10330C thermoelectrically-cooled near-infrared PMT. High-resolution laser transmission measurements were taken by cooling the sample inside a Bluefors LD250 dilution fridge to $\sim$3 K and then sweeping a Toptica CTL1500 single-frequency laser with a linewidth of the order of 100~kHz across the absorption transition.

\section{Results}\label{results}
 In this work, we adopt the labelling scheme of Dieke \cite{Dieke:63}. Thus the ground state multiplet is labelled Z with the first excited multiplet labelled Y and so on. Numerical subscripts are used to denote individual crystal-field states within a given multiplet.
 
 \subsection{Absorption and Laser Spectroscopy}

\begin{figure}[!tb]
	\centering
		\includegraphics[width=0.5\columnwidth]{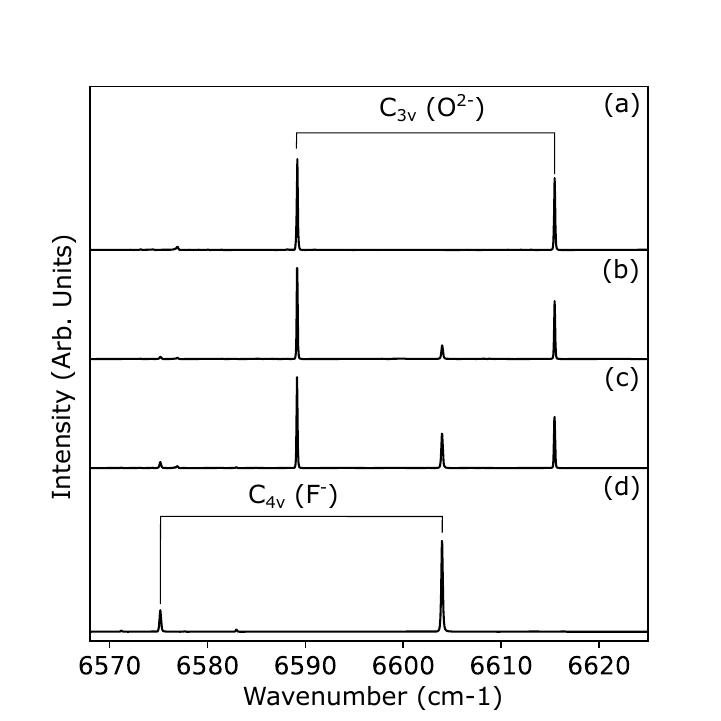}
	  \caption{10~K absorption spectra of oxygenated and as-grown CaF$_{2}$:0.01\%Er$^{3+}$. (a) 5-hour annealing at 1100 $^{\circ}$C, (b) 24-hour annealing at 850 $^{\circ}$C, (c) 5-hour annealing at 850 $^{\circ}$C, and (d) as grown.  }\label{Fig:Y12_Comparison}
\end{figure}

 The effect of oxygenation is demonstrated by the 10~K infrared absorption spectra in Figure \ref{Fig:Y12_Comparison}, which shows the lowest-energy transitions in the  $^4$I$_{15/2}$ $\rightarrow$ $^4$I$_{13/2}$ spectra of non-annealed and annealed samples.
 The spectrum of the as-grown sample is dominated by transitions associated with the C$_{\rm 4v}$(F$^{-}$) centre.
 In the oxygenated samples, the transitions associated with the C$_{\rm 4v}$(F$^{-}$) centre are reduced and the  6589.6 and 6615.5 cm$^{-1}$  transitions associated with the C$_{\rm 3v}$(O$^{2-}$) centre increase.
The  population of this oxygen charge-compensated centre correlates with both annealing time and temperature, as expected due to the oxygen diffusion process \cite{phillips}. The most heavily oxygenated sample exhibits complete conversion to the C$_{\rm 3v}$(O$^{2-}$) centre. 
 
\begin{figure*}[!tb]
	\centering
		\includegraphics[width=\textwidth]{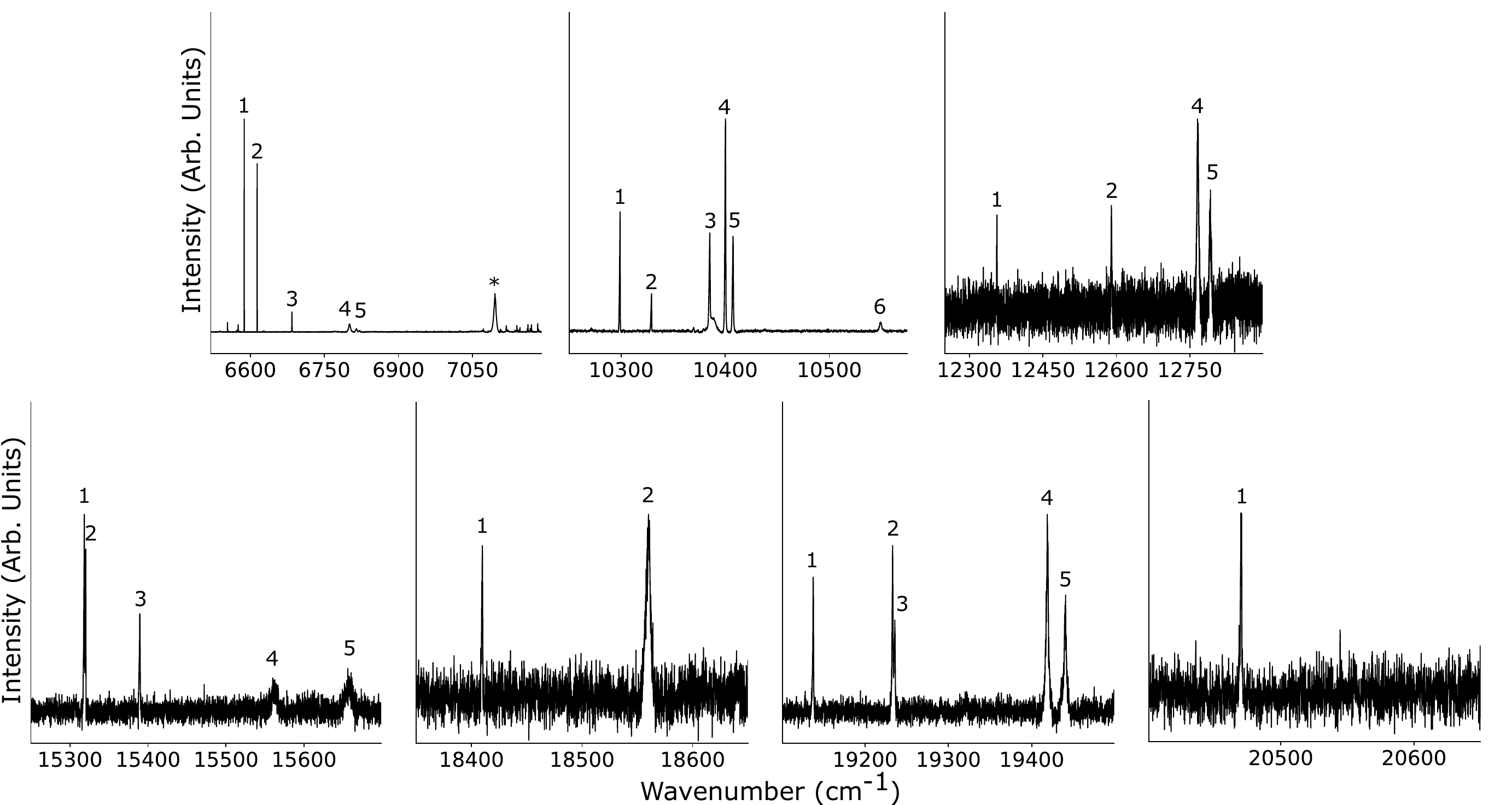}
                \caption{10 K absorption spectra of CaF$_{2}$:0.01\%Er$^{3+}$ annealed at $1100 ^{\circ}C$ for 5 hours showing the ground state $^{4}$I$_{15/2}$Z$_1$ $\rightarrow$ (a) $^{4}$I$_{13/2}$, (b) $^{4}$I$_{11/2}$, (c) $^{4}$I$_{9/2}$, (d) $^{4}$F$_{9/2}$, (e) $^{4}$S$_{3/2}$, (f) $^{2}$H$_{11/2}$ and (g) $^{4}$F$_{7/2}$ excited state transitions. Peaks assigned to the C$_{\rm 3v}$(O$^{2-}$) centre are labelled, with the numbers within each multiplet corresponding to those in Table \ref{tab:energylevels}.
 The absorption peak at $\sim$7095 cm$^{-1}$ marked with an asterisk is attributed to an impurity. 
 \label{absorption}}
\end{figure*}

\highlight{
Figure \ref{absorption} presents 
10~K absorption spectra. 28  of the 35 expected excited-state crystal-field levels from the $^{4}$I$_{13/2}$ multiplet at approximately 6700 cm$^{-1}$ through to the $^{4}$F$_{7/2}$ multiplet at approximately 20600 cm$^{-1}$ are labelled. The $^{4}$I$_{9/2}$B$_{3}$ level is not visible in Figure \ref{absorption}, but was observed in other scans.
The absorption peak at $\sim$7095 cm$^{-1}$ marked with an asterisk is present in annealed samples of other rare-earth doped CaF$_2$ crystals and therefore can not be attributed to any Er$^{3+}$ transitions.
The $^{4}$F$_{7/2}$G$_{2,3}$ levels were assigned via laser site-selective excitation spectroscopy.
All 8 of the $^4$I$_{15/2}$ levels were determined by laser site-selective fluorescence measurements (see Figure \ref {fig:intensities} for a selection of these measurements). In total, 39 of the possible 43 crystal-field levels below 21,000\,cm$^{-1}$ were assigned. These are listed in Table \ref{tab:energylevels}. 
}

\begin{figure}[!tb]
	\centering
		\includegraphics[width=0.5\columnwidth]{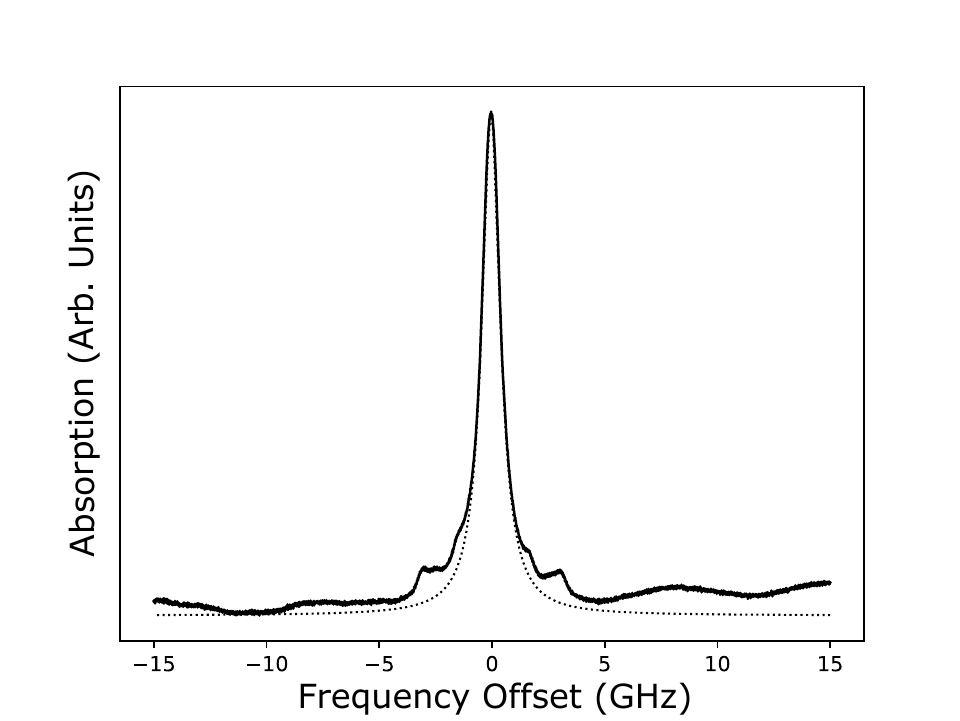}
                \caption{3~K absorption spectra of the $^4$I$_{15/2}$ Z$_{1}$$\rightarrow$ $^{4}$I$_{13/2}$ Y$_1$ transition in CaF$_{2}$:0.01\%Er$^{3+}$ centred at 6589.6 cm$^{-1}$. The solid black line is the experimental data and the dashed black line is  a Lorentzian fit to the central peak, giving a linewidth (FWHM) of 0.99\,GHz. 
}\label{high res line}
\end{figure}

Figure \ref{high res line} shows a high-resolution measurement of the $^4$I$_{15/2}$ Z$_{1}$$\rightarrow$ $^{4}$I$_{13/2}$Y$_1$ transition at 6589.6 cm$^{-1}$  using a transmission sweep of a 100 kHz linewidth external cavity diode laser at sample temperature of 3~K. The inhomogeneous linewidth was was measured to be 0.99 GHz. There is visible sideband structure around the central absorption peak, which is attributed to hyperfine splittings of the $^{167}$Er isotope, which has a nuclear spin of 7/2 and a natural abundance of 22.9\%.

\subsection{Crystal-Field Analysis}

Table \ref{tab:energylevels} shows the crystal-field fit and Table \ref{tab:cfparams} the fitted parameters.
The starting point for the fit was the Er$^{3+}$ free-ion parameters from Ref.\ \cite{Carnall} and the  crystal-field parameters from Ref.\ \cite{Cockroft90}
\highlight{
(Table \ref{tab:calcparams}(a)).
}
The computer code described in Ref.\ \cite{sebastian} was used to fit 34 experimentally-determined energy levels and the two-ground state magnetic splitting factors, for fields parallel and perpendicular to the C$_3$ axis of the site. The latter were determined by electron paramagnetic resonance \cite{ranon1963electron}. The calculated and experimental ground state $^4$I$_{15/2}$ Z$_{1}$ magnetic splitting factors are given Table \ref{tab:ZYcalculations}.  In the fit the $\chi^2$ of the two magnetic splittings was weighted higher than the 34 energy levels by a factor of 3,000. Uncertainties were estimated using the Monte-Carlo technique described in  Ref.\ \cite{sebastian}. 
The crystal-field levels of the $^{2}$H$_{11/2}$ multiplet were omitted from the fit. This  multiplet is  known to be affected by correclation crystal-field effects \cite{ReidCor}. It is possible to add extra parameters to account for these effects, but in the current study the data were not complete enough for this to be practical.
\highlight{
The fit gave parameters quite similar to those of Ref.\ \cite{Cockroft90}, even though the latter were obtained using a much smaller data set. 
}

The large value of the $B_{0}^{k}$ parameters seen in Table \ref{tab:cfparams} is expected due to the potential felt by the central Er$^{3+}$ ion from the substitutional O$^{2-}$. These axial parameters are also predicted to be very large by the point-charge calculation, Table \ref{tab:calcparams}(a).  The signs of the fitted parameters agree with the point-charge calculation, though, as noted above, the magnitudes are not accurate. The fitted ratio $B^6_6/B^6_3$ is smaller than the point-charge value, which may indicate that there is  some distortion of the cube surrounding the Er$^{3+}$ ion.

\begin{table*}[!tb]
  \caption{Experimental and calculated energy levels and oscillator strengths for the C$_{\rm 3v}$ (O$^{2-}$) centre of Er$^{3+}$ in CaF$_{2}$. Data marked with an asterisk were excluded from the fit. Experimental levels are determined with an uncertainty of $\pm$1 cm$^{-1}$ from laser spectroscopy and $\pm$0.1 cm$^{-1}$ from FTIR absorption spectroscopy.
\label{tab:energylevels} }

\footnotesize
\begin{tabular*}{\tblwidth}{@{}LLLLLLLRR@{}}
\toprule
       \multicolumn{2}{c}{Label}&\multicolumn{2}{c}{Energy (in cm$^{-1}$)} & \multicolumn{3}{c}{Osc. Str. (in $10^{-10}$)} \\ 
       \midrule
  Multiplet & State and Symmetry & Exp & Calc & Exp & Calc \\
  \midrule

$^4$I$_{15/2}$ &  Z$_1$ ($\gamma _{4}$)    &                        0.0  &      2.8& -             & -          \\     
~              &  Z$_2$ ($\gamma _{5,6}$)  &                         51  &     50.9& -             & -          \\ 
~              &  Z$_3$ ($\gamma _{4}$)    &                        100  &    105.5& -             & -          \\
~              &  Z$_4$ ($\gamma _{4}$)    &                        200  &    198.2& -             & -          \\
~              &  Z$_5$ ($\gamma _{4}$)    &                        429  &    419.7& -             & -          \\
~              &  Z$_6$ ($\gamma _{5,6}$)  &                        444  &    441.0& -             & -          \\
~              &  Z$_7$ ($\gamma _{4}$)    &                        460  &    455.4& -             & -          \\
~              &  Z$_8$ ($\gamma _{5,6}$)  &                        750  &    754.4& -             & -          \\ 
$^4$I$_{13/2}$ &  Y$_1$ ($\gamma _{4}$)    &                     6589.6  &   6587.2&  2407*	   & 2522  \\      
~              &  Y$_2$ ($\gamma _{5,6}$)  &                     6615.5  &   6615.3&  1551 	   &  778  \\
~              &  Y$_3$ ($\gamma _{4}$)    &                     6685.8  &   6691.0&   293 	   &  460  \\
~              &  Y$_4$ ($\gamma _{4}$)    &                     6802.7  &   6801.6&  1435 	   &  409  \\
~              &  Y$_5$ ($\gamma _{5,6}$)  &                     6822.9  &   6827.2&   350* 	   &  766  \\     
~              &  Y$_6$ ($\gamma _{4}$)    & -                           &   6844.7&     - 	   & 1193  \\
~              &  Y$_7$ ($\gamma _{4}$)    & -                           &   7086.5&     - 	   &  202  \\
$^4$I$_{11/2}$ &  A$_1$ ($\gamma _{4}$)    &                    10301.4  &  10290.4&   574 	   &  851  \\
~              &  A$_2$ ($\gamma _{5,6}$)  &                    10331.6  &  10323.3&   123 	   &  275  \\
~              &  A$_3$ ($\gamma _{4}$)    &                    10387.7  &  10384.7&   426 	   &  196  \\
~              &  A$_4$ ($\gamma _{4}$)    &                    10402.8  &  10410.9&  1120 	   & 1328  \\
~              &  A$_5$ ($\gamma _{5,6}$)  &                    10410.1  &  10420.3&   509 	   &  150  \\
~              &  A$_6$ ($\gamma _{4}$)    &                    10549.2  &  10587.4&    25* 	   &   47  \\      
$^4$I$_{9/2}$  &  B$_1$ ($\gamma _{4}$)    &                    12359.3  &  12348.0&    57 	   &  160  \\
~              &  B$_2$ ($\gamma _{4}$)    &                    12593.3  &  12605.1&    69 	   &   52  \\
~              &  B$_3$ ($\gamma _{5,6}$)  &                    12616.6  &  12622.0&     - 	   &  157  \\
~              &  B$_4$ ($\gamma _{4}$)    &                    12769.5  &  12767.0&   677 	   &  531  \\
~              &  B$_5$ ($\gamma _{5,6}$)  &                    12795.0  &  12789.6&   440* 	   &  339  \\       
$^4$F$_{9/2}$  &  D$_1$ ($\gamma _{5,6}$)  &                     15322.0  &  15331.2&  943 	   & 1320  \\
~              &  D$_2$ ($\gamma _{4}$)    &                    15323.8  &  15336.1&   840 	   & 1155  \\
~              &  D$_3$ ($\gamma _{4}$)    &                    15393.3  &  15393.2&   360 	   &  586  \\
~              &  D$_4$ ($\gamma _{5,6}$)  &                    15565.9  &  15564.4&  1045 	   &  634  \\
~              &  D$_5$ ($\gamma _{4}$)    &                    15651.5  &  15645.6&  1400* 	   & 1021  \\       
$^4$S$_{3/2}$  &  E$_1$ ($\gamma _{5,6}$)  &                    18414.2  &  18426.6&  1992 	   & 1901  \\
~              &  E$_2$ ($\gamma _{4}$)    &                    18564.6  &  18553.5&  7011 	   & 4071  \\
$^2$H$_{11/2}$ &  F$_1$ ($\gamma _{4}$)    & 19142.3$\ast$               &  19251.2&  7486 	   & 8524  \\
~              &  F$_2$ ($\gamma _{4}$)    & 19237.8$\ast$               &  19259.6& 10041 	   &17546  \\
~              &  F$_3$ ($\gamma _{5,6}$)  & 19240.4$\ast$               &  19264.1&  5002 	   & 7623  \\
~              &  F$_4$ ($\gamma _{4}$)    & 19423.8$\ast$               &  19440.4& 23447 	   & 8059  \\
~              &  F$_5$ ($\gamma _{4}$)    & 19445.1$\ast$               &  19447.1& 11493 	   &15426  \\
~              &  F$_6$ ($\gamma _{5,6}$)  & -                           &  19466.9&     - 	   &15409  \\
$^4$F$_{7/2}$  &  G$_1$ ($\gamma _{4}$)    &                    20475.4  &  20462.8&  9720 	   & 5153  \\
~              &  G$_2$ ($\gamma _{4}$)    &                      20623  &  20619.5&     - 	   & 1022  \\
~              &  G$_3$ ($\gamma _{5,6}$)  &                      20746  &  20752.3&     - 	   & 1213  \\
~              &  G$_4$ ($\gamma _{4}$)    & -                           &  20758.1&     -         & 5270  \\

\bottomrule
\end{tabular*}
\end{table*}

\begin{table}[!tb]
  \caption{Free-ion and crystal-field parameters (in cm$^{-1}$) of the C$_{\rm 3v}$(O$^{2-}$) centre in Er$^{3+}$ doped CaF$_{2}$. The values in square brackets were held constant. 
    $M^2$ and $M^4$ were set to $0.558 M^0$ and $0.377 M^0$ respectively,
    $P^4$ and $P^6$ were set to $0.75 P^2$ and $0.50 P^2$ respectively. 
  }
\label{tab:cfparams}
\footnotesize
\begin{tabular}{ll}
\toprule

        Parameter & Fit \\ 
        \midrule
        $E_{avg}$ & 35728$\pm$14 \\
        $F^2$ & 96493$\pm$85 \\
        $F^4$ & 68015$\pm$89 \\
        $F^6$ & 53967$\pm$48 \\
        $\alpha$ & [17.8] \\
        $\beta$ & [-582] \\
        $\gamma$ & [1800] \\
        $T^2$ & [400] \\
        $T^3$ & [43] \\
        $T^4$ & [73] \\
        $T^6$ & [-271] \\
        $T^7$ & [308] \\
        $T^8$ & [299] \\
        $M^{0}$ & [3.86] \\
        $P^{2}$ & [594] \\
        $\zeta$ & 2362$\pm$8 \\
        $B_{0}^{2}$ & 1268$\pm$45 \\
        $B_{0}^{4}$ & 1760$\pm$99 \\
        $B_{3}^{4}$ & 1233$\pm$32 \\
        $B_{0}^{6}$ & 792$\pm$79 \\
        $B_{3}^{6}$ & -658$\pm$47 \\
        $B_{6}^{6}$ & 418$\pm$46 \\
\bottomrule
\end{tabular}
\end{table}


\begin{table}[!tb]
  \caption{Data and calculations for the C$_{\rm 3v}$(O$^{2-}$) centre in Er$^{3+}$ doped CaF$_{2}$.
    Energies (in cm$^{-1}$) and g-values for the Z and Y multiplets. Einstein $A$ coefficients (in s$^{-1}$) for emission from  Y$_1$  to the Z multiplet.
    The energies in parentheses are the calculated values for states for which there is no  experimental data.
\label{tab:ZYcalculations}
  }
  \footnotesize
  \begin{tabular}{@{}rllllll@{}}
    \toprule
    State & Energy &\multicolumn{2}{c}{$g_{\parallel}$}&\multicolumn{2}{c}{$g_{\perp}$}&{$A$} \\
    \midrule
          & & Exp & Calc & Exp & Calc & Calc \\
     \hline
        $^4$I$_{15/2}$~Z$_1$ & 0.0 & 8.84 & 8.95 & 2.21 & 2.34 &    14.97 \\ 
        Z$_2$ & 51 & - & 0.00 & - & 6.86                      &      9.66\\
        Z$_3$ & 100 & - & 1.26 & - & 11.46                    &      1.06\\
        Z$_4$ & 200 & - & 1.09 & - & 14.95                    &      0.78\\
        Z$_5$ & 429 & - & 6.21 & - & 3.73                     &      3.00  \\
        Z$_6$ & 444 & - & 0.00 & - & 6.78                     &      1.68\\
        Z$_7$ & 460 & - & 6.68 & - & 3.93                     &      3.8 \\
         Z$_8$ & 750 & - & 0.00 & - & 17.69                   &      0.26\\
    
        $^4$I$_{13/2}$~Y$_1$ & 6589.6 & - & 7.15 & - & 2.28 & ~ \\
        Y$_2$ & 6615.5 & - & 0.00 & - & 7.08 & ~ \\
        Y$_3$ & 6685.8 & - & 0.65 & - & 9.76 & ~ \\
        Y$_4$ & 6802.7 & - & 5.10 & - & 0.48 & ~ \\
        Y$_5$ & 6822.9 & - & 0.00 & - & 5.70 & ~ \\
        Y$_6$ & (6844.7) & - & 6.39 & - & 1.04 & ~ \\
        Y$_7$ & (7086.5) & - & 0.01 & - & 14.31 \\
        \bottomrule
\end{tabular}
\end{table}

\clearpage 
\subsection{Transition Intensities}

\subsubsection{Intensity Fit}

\begin{table}[!tb]
  \caption{Fitted electric-dipole transition-intensity parameters (in $10^{-12}$\,cm) for the C$_{\rm 3v}$(O$^{2-}$) centre in CaF$_{2}$:Er$^{3+}$. } \label{Intensity_params}
\footnotesize  
\begin{tabular}{ll}
\toprule
          Parameter & Fit              \\
 
  \midrule
        A$^2_{10}$  & 486  $ \pm   61 $   \\
        A$^2_{30}$  & 110  $ \pm   93 $   \\
        A$^4_{30}$  &-189  $ \pm   33 $   \\
        A$^4_{50}$  &-133  $ \pm   38 $   \\
        A$^6_{50}$  &  74  $ \pm   44 $   \\
        A$^6_{70}$  &-197  $ \pm   38 $   \\
\bottomrule
\end{tabular}
\end{table}

Experimental and calculated oscillator strengths are shown in Table \ref{tab:energylevels}. Intensity parameters are shown in Table \ref{Intensity_params}. Oscillator strengths were estimated from the absorption spectrum. The exact concentration of C$_{\rm 3v}$(O$^{2-}$) centres is unknown, due to the multi-site nature of the crystal.
\highlight{
However, the calculations indicated that the oscillator strength of the Z$_1$ $\rightarrow $Y$_1$ transition was almost entirely magnetic-dipole. Therefore, this transition could be used to calibrate the absorption measurements. The experimental absorption data were scaled so that the oscillator strength of the Z$_1$ $\rightarrow $Y$_1$  transition matched the calculated magnetic-dipole oscillator strength. This transition was omitted from the fit, along with a number of other transitions that could not be measured accurately. Other absorption transitions with significant magnetic-dipole character were  Z$_1$ $\rightarrow $Y$_2$ and Z$_1$ $\rightarrow $Y$_4$. Emission from  Y$_1$ to  Z$_1$, Z$_2$, and Z$_4$ was also predominately magnetic-dipole and emission from  Y$_1$ to  Z$_8$ had significant magnetic-dipole character. Comparison of the calculated oscillator strengths with absorption coefficients suggest that  the majority of the Er$^{3+}$ ions are in C$_{\rm 3v}$(O$^{2-}$) centres.
}



The fit was obtained by minimizing the sum of $|(\text{experimental} - \text{calculated})/(\text{experimental} + \text{
\highlight{
calculated}})|^2$. This approach prevents the fit from being dominated by the more intense transitions. See, for example, Ref.\ \cite{BuJaRiRe94}. We also carried out fits minimizing the sum of $|\text{experimental} - \text{calculated}|^2$. In that case, the sign of $A^2_{10}$ became negative. From Table \ref{Intensity_params}, we can see that this parameter has a very large uncertainty. The fitted parameters have signs consistent with the calculations in Table \ref{tab:calcparams}, apart from $A^6_{70}$. 
As expected, the parameters are very large. $A^2_{10}$ is similar to the Eu$^{3+}$:DBM system studied in Ref.\ \cite{DaReRi84}. Yttrium aluminium garnet also has O$^{2-}$ ligands. In the D$_2$ site symmetry of that host there are no $A^2_{10}$ parameters. However, fitted intensity parameters for Nd$^{3+}$ are of similar magnitude to those obtained here \cite{BuJaRiRe94}


\subsubsection{Simulated emission spectra}

\begin{figure*}[!tb]
	\centering
        \includegraphics[width=0.49\textwidth]{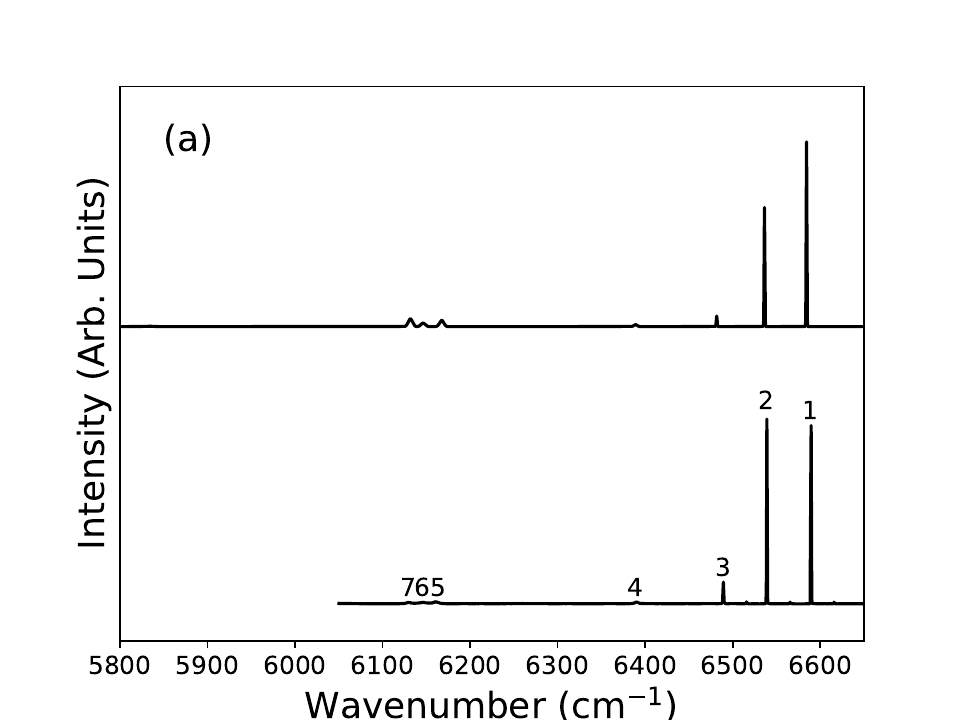}
        \includegraphics[width=0.49\textwidth]{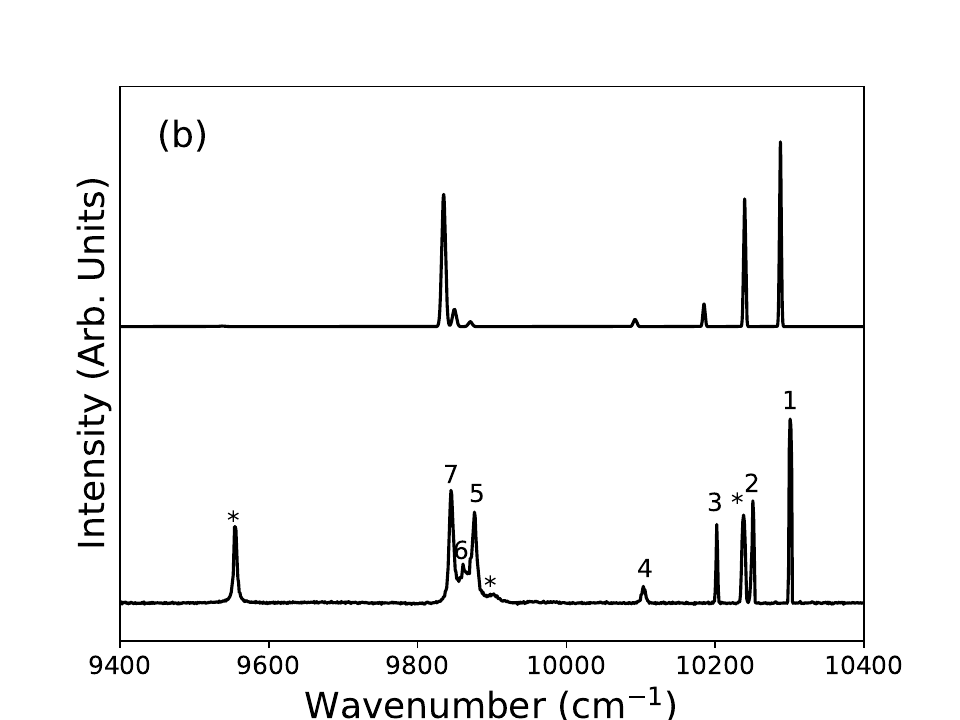}
                \caption{10~K calculated (top) and experimental (bottom) fluorescence spectra for the (a) $^{4}$I$_{13/2}$$\rightarrow$$^{4}$I$_{15/2}$ and (b) $^{4}$I$_{11/2}$$\rightarrow$$^{4}$I$_{15/2}$ transitions of the C$_{\rm 3v}$(O$^{2-}$) centre in CaF$_2$:0.01\%Er$^{3+}$. The transition labelled with asterisks are overlapping spectra from different multiplets. 
                \label{fig:intensities}}
\end{figure*}

Figure \ref{fig:intensities} presents simulated emission spectra using the  intensity parameters from the weighted fit.  The theoretical $A$ coefficients for each line were convolved with a Gaussian with the same  width as  the experimental spectral lines.  Note that the signal for the the Y$_1$ $\rightarrow$ Z$_{5,6,7}$ transitions are weak, and the  Y$_1$ $\rightarrow$ Z$_{8}$ transition is not visible at all, due to a steep decline in the Hamamatsu H10330C near-infrared PMT's sensitivity at these energies.

\subsubsection{$^4$I$_{13/2}$ excited state lifetime}

\begin{figure}[!tb]
    \centering
    \includegraphics[width=0.5\columnwidth]{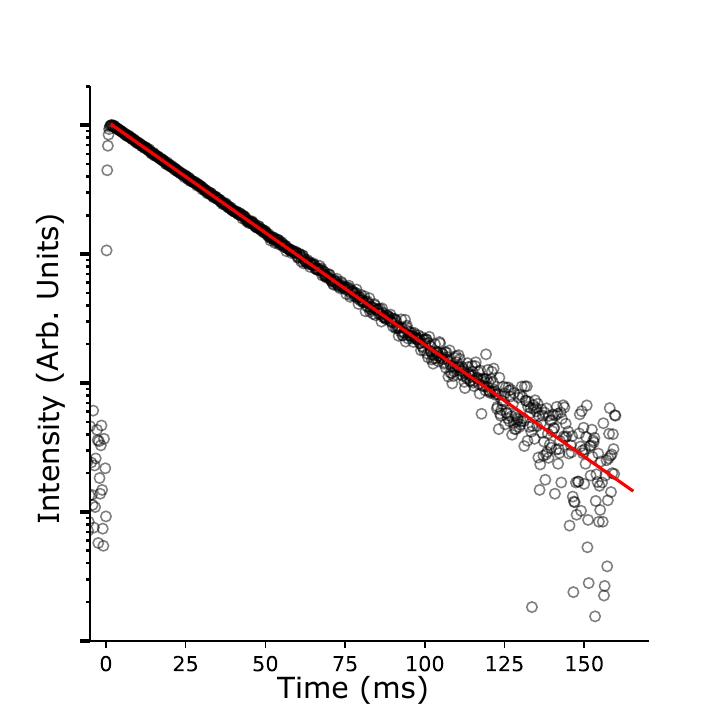}
    \caption{10 K fluorescence transient measured monitoring the $^4$I$_{13/2}$(Y$_1$) $\rightarrow$ $^4$I$_{15/2}$(Z$_1$) transition of the C$_{\rm 3v}$(O$^{2-}$) centre in CaF$_2$:0.01\%Er$^{3+}$. The solid red line represents the fit of a single exponential function to the experimental data, giving a lifetime $\tau = 25$\,ms.
    \label{fig:Lifetime} }
  \end{figure}

  Figure \ref{fig:Lifetime} shows 10 K fluorescence transient measured whilst monitoring the Y$_1$ $\rightarrow$ Z$_1$ transition at 6589.6 cm$^{-1}$. The decay is single exponential with a lifetime of 25.0 ms.
\highlight{
The decay time varied only slowly with  temperature and was still more than 20 ms at 120 K. A low non-radiative rate is expected at cyrogenic temperatures, due to the large energy gap between the $^{4}$I$_{13/2}$ and
$^{4}$I$_{15/2}$ multiplets, which is well in excess of ten times the maximum phonon energy.

Given the low non-radiative relaxation rate, the calculated Einstein $A$ coefficients presented in Table \ref{tab:ZYcalculations} should reproduce the experimental fluorescence lifetime.
The  calculation gives an $^4$I$_{13/2}$ Y$_1$ excited state lifetime of 28\,ms. The magnetic-dipole contribution is larger than the electric dipole contribution. The magnetic-dipole contribution alone gives a lifetime of 41\,ms. A change in the magnitude of the $A^\lambda_{tp}$ parameters by about 20\% would bring the calculation into agreement with experiment. This is well within the parameter uncertainties. 

The  lifetime of $^4$I$_{13/2}$(Y$_1$) is often 10\,ms or more, due to the low frequency of the radiation and the $\omega^3$ dependence of the $A$ coefficients (Eq.\,(\ref{eq:radlifetime}). 
Given the large electric-dipole intensity parameters for this site, it is surprising that the radiative lifetime is so long, longer than some Er$^{3+}$ sites with purely magnetic-dipole Y to Z transitions \cite{stevenson2022erbium}. However, at low temperatures the lifetime depends only on the transitions from Y$_1$, so the lifetime is determined by a complex interplay between the wavefunction for  Y$_1$, the wavefunctions for the Z multiplet, and the intensity parameters. Futhermore, the refractive index of CaF$_2$ is quite low (1.43) compared to some of the materials studied in Ref.\ \cite{stevenson2022erbium}. 
}

\section{Conclusions}\label{conclusion}

Crystal-field modelling is a powerful tool to understand and leverage the electronic structure of rare-earth ion crystals for quantum information networking hardware. The properties of trivalent erbium are particularly important because they present the opportunity to capitalise on low-loss fiber-optic infrastructure for quantum applications. Toward the aim of deepening the understanding of erbium systems, we have presented a detailed spectroscopic analysis of the C$_{\rm 3v}$(O$^{2-}$) centre in CaF$_2$ crystals doped with Er$^{3+}$.
\highlight{
A parametrized crystal-field Hamiltonian 
}
was fitted to 34 energy levels and the two ground-state magnetic splitting factors, and electric-dipole intensity parameters were fitted to absorption  oscillator strengths. Simulated spectra, including electric-dipole and magnetic-dipole contributions, gave a  good approximation to the data and calculated transition rates reproduce the surprisingly long 
25~ms  $^{4}$I$_{13/2}$ lifetime. Though the electric-dipole parameters are large, this particular lifetime is dominated by magnetic-dipole contributions. Despite the long lifetime, the oscillator strength of this site is similar to other erbium sites currently under investigation for single Er$^{3+}$ technologies. This motivates future work to investigate the coherence properties of this site and to what extent the crystal field can be engineered to enhance the oscillator strength of erbium towards the optimized site for quantum information science.  


\section*{Acknowledgements}
This research was supported by the Australian Research Council Centre of Excellence for Engineered Quantum Systems (EQUS, CE170100009) and the Discovery Project (Grant DP210101784). The authors would also like to acknowledge the support of a Sydney Nano – MacDiarmid Institute – Dodd Walls Centre Joint research award.




\begin{thebibliography}{10}

\bibitem{thiel2011rare}
C.W. Thiel, T.~B{\"o}ttger, and R.L. Cone.
\newblock Rare-earth-doped materials for applications in quantum information
  storage and signal processing.
\newblock {\em J. Lumin.}, 131(3):353--361, 2011.

\bibitem{goldner2015rare}
P.~Goldner, A.~Ferrier, and O.~Guillot-No{\"e}l.
\newblock Rare earth-doped crystals for quantum information processing.
\newblock In {\em Handbook on the Physics and Chemistry of Rare Earths},
  volume~46, pages 1--78. Elsevier, 2015.

\bibitem{Becher2023}
C.~Becher, W.~Gao, S.~Kar, C.~D. Marciniak, T.~Monz, J.~G. Bartholomew,
  P.~Goldner, H.~Loh, E.~Marcellina, K.~E.~J. Goh, T.~S. Koh, B.~Weber, Z.~Mu,
  J.-Y. Tsai, Q.~Yan, T.~Huber-Loyola, S.~H{\"{o}}fling, S.~Gyger,
  S.~Steinhauer, and V.~Zwiller.
\newblock {Rare Earth Ions for Optical Quantum Technologies, Section 3, 2023
  roadmap for materials for quantum technologies}.
\newblock {\em Materials for Quantum Technology}, 3(1):012501, mar 2023.

\bibitem{ruskuc2024}
A.~Ruskuc, C.-J. Wu, E.~Green, S.~L.~N. Hermans, J.~Choi, and A.~Faraon.
\newblock Scalable multipartite entanglement of remote rare-earth ion qubits.
\newblock (arXiv:2402.16224), February 2024.

\bibitem{uysal2024}
Mehmet~T. Uysal, {\L}ukasz Dusanowski, Haitong Xu, Sebastian~P. Horvath, Salim
  Ourari, Robert~J. Cava, Nathalie~P. {de Leon}, and Jeff~D. Thompson.
\newblock Spin-photon entanglement of a single
  {{{Er}}\${\textasciicircum}\{3+\}\$} ion in the telecom band.
\newblock (arXiv:2406.06515), June 2024.

\bibitem{raha2021}
M.~Raha.
\newblock {\em A Telecom-Compatible Quantum Memory in the Solid-State: Single
  Erbium Ions Coupled to Silicon Nanophotonic Circuits}.
\newblock PhD thesis, Princeton University, 2021.

\bibitem{stevenson2022erbium}
P.~Stevenson, C.M. Phenicie, I.~Gray, S.P. Horvath, S.~Welinski, A.M. Ferrenti,
  A.~Ferrier, P.~Goldner, S.~Das, R.~Ramesh, et~al.
\newblock Erbium-implanted materials for quantum communication applications.
\newblock {\em Phys. Rev. B.}, 105(22):224106, 2022.

\bibitem{sebastian}
S.H. Horvath, J.V. Rakonjac, Y.-H. Chen, J.J. Longdell, P.~Goldner, J.-P.R.
  Wells, and M.F. Reid.
\newblock Extending phenomenological crystal-field methods to {C$_{1}$}
  point-group symmetry: Characterization of the optically excited hyperfine
  structure of {$^{167}$Er$^{3+}$:Y$_{2}$SiO$_{5}$}.
\newblock {\em Phys. Rev. Lett.}, 123:057401, 2019.

\bibitem{transfer}
N.L. Jobbitt, S.J. Patchett, Y.~Alizadeh, M.F. Reid, J.-P.R. Wells, S.P.
  Horvath, J.J. Longdell, A.~Ferrier, and P.~Goldner.
\newblock Transferability of crystal-field parameters for rare-earth ions in
  {Y$_{2}$SiO$_{5}$} tested by {Z}eeman spectroscopy.
\newblock {\em Phys. Sol. Stat.}, 61:780--784, 2019.

\bibitem{zhou_2020}
X.~Zhou, H.~Liu, Z.~He, B.~Chen, and J.~Wu.
\newblock Investigation of the electronic structure and optical, epr, and odmr
  spectroscopic properties for $^{171}\mathrm{Yb}^{3+}$-doped
  $\mathrm{Y}_{2}\mathrm{SiO}_{5}$ crystal: A combined theoretical approach.
\newblock {\em Inorg. Chem.}, 59(18):13144--13152, 2020.
\newblock PMID: 32865403.

\bibitem{yasharce}
Y.~Alizadeh, J.L.B Martin, M.F. Reid, and J.-P.R. Wells.
\newblock Intra- and inter-configurational electronic transitions of
  {Ce$^{3+}$} doped {Y$_{2}$SiO$_{5}$}: Spectroscopy and crystal-field
  analysis.
\newblock {\em Opt. Mat.}, 117:111114, 2021.

\bibitem{yasharnd}
Y.~Alizadeh, J.-P.R. Wells, M.F. Reid, A.~Ferrier, and P.~Goldner.
\newblock {Z}eeman spectroscopy and crystal-field analysis of low symmetry
  centres in {Nd$^{3+}$} doped {Y$_{2}$SiO$_{5}$}.
\newblock {\em J. Phys.: Condens. Matter}, 35:305502, 2023.

\bibitem{nicksm}
N.L. Jobbitt, J.-P.R. Wells, and M.F. Reid.
\newblock {Z}eeman and laser site selective spectroscopy of {C$_{1}$} point
  group symmetry {Sm$^{3+}$} centres in {Y$_{2}$SiO$_{5}$}: a parametrized
  crystal-field analysis for the {4$f^{5}$} configuration.
\newblock {\em J. Phys.: Condens. Matter}, 34:325502, 2022.

\bibitem{nicker}
N.L. Jobbitt, J.-P.R. Wells, M.F. Reid, S.P. Horvath, P.~Goldner, and
  A.~Ferrier.
\newblock Prediction of optical polarization and high-field hyperfine structure
  via a parametrized crystal-field model for low-symmetry centres in
  {Er$^{3+}$} doped {Y$_{2}$SiO$_{5}$}.
\newblock {\em Phys. Rev. B.}, 104:155121, 2021.

\bibitem{McAuslan2009}
D.~L. McAuslan, J.~J. Longdell, and M.~J. Sellars.
\newblock {Strong-coupling cavity QED using rare-earth-metal-ion dopants in
  monolithic resonators: What you can do with a weak oscillator}.
\newblock {\em Physical Review A}, 80(6):1--9, 2009.

\bibitem{dibos2018}
A.~M. Dibos, M.~Raha, C.~M. Phenicie, and J.~D. Thompson.
\newblock Atomic {{Source}} of {{Single Photons}} in the {{Telecom Band}}.
\newblock {\em Phys Rev Lett}, 120(24):243601, June 2018.

\bibitem{Xie2021}
T.~Xie, J.~Rochman, J.~G. Bartholomew, A.~Ruskuc, J.~M. Kindem, I.~Craiciu,
  C.~W. Thiel, R.~L. Cone, and A.~Faraon.
\newblock {Characterization of {Er$^{3+}$:YVO$_4$} for microwave to optical
  transduction}.
\newblock {\em Physical Review B}, 104(5):54111, 2021.

\bibitem{ourari2023}
S.~Ourari, {\L}~Dusanowski, S.~P. Horvath, M.~T. Uysal, C.~M. Phenicie,
  P.~Stevenson, M.~Raha, S.~Chen, R.~J. Cava, N.~P. de~Leon, and J.~D.
  Thompson.
\newblock Indistinguishable telecom band photons from a single {{Er}} ion in
  the solid state.
\newblock {\em Nature}, 620:977--981, August 2023.

\bibitem{wells}
J.-P.R. Wells, G.D. Jones, M.F. Reid, M.N. Popova, and E.P. Chukalina.
\newblock Hyperfine patterns of infrared absorption lines of {Ho$^{3+}$}
  {C$_{\rm 4v}$} centres in {CaF$_{2}$}.
\newblock {\em Mol. Phys.}, 102:1367--1376, 2004.

\bibitem{hozeeman}
K.M. Smith, M.F. Reid, and J.-P.R. Wells.
\newblock {Z}eeman-hyperfine measurements of a pseudo-degenerate quadruplet in
  {CaF$_{2}$:Ho$^{3+}$}.
\newblock {\em Opt. Spectrosc.}, 131:460--465, 2023.

\bibitem{tallant1975selective}
D.R. Tallant and J.C. Wright.
\newblock Selective laser excitation of charge compensated sites in
  {CaF$_2$:Er$^{3+}$}.
\newblock {\em J. Chem. Phys.}, 63(5):2074--2085, 1975.

\bibitem{gustafson1979trace}
F.J. Gustafson and J.C. Wright.
\newblock Trace analysis of lanthanides by laser excitation of precipitates.
\newblock {\em Anal. Chem.}, 51(11):1762--1774, 1979.

\bibitem{ranon1963electron}
U.~Ranon and W.~Low.
\newblock Electron spin resonance of {Er$^{3+}$} in {CaF$_{2}$}.
\newblock {\em Phys. Rev.}, 132(4):1609, 1963.

\bibitem{Cockroft90}
N.J. Cockroft, G.D. Jones, and R.W.G. Syme.
\newblock Site‐selective laser spectroscopy of deuterated
  {SrF$_2$:Er$^{3+}$}.
\newblock {\em J. Chem. Phys.}, 92(4):2166--2177, 1990.

\bibitem{chehaidar2000interstitial}
A.~Chehaidar and L.~Hirsch.
\newblock Interstitial-fluoride and substitutional-oxygen charge compensations
  of {Er$^{3+}$} doped in {CaF$_2$} crystal.
\newblock {\em Eur. Phys. J.- Appl. Phys.}, 12(2):79--83, 2000.

\bibitem{smith2022complete}
K.M. Smith, M.F. Reid, M.J. Sellars, and R.L. Ahlefeldt.
\newblock Complete crystal-field calculation of {Z}eeman hyperfine splittings
  in europium.
\newblock {\em Phys. Rev. B.}, 105(12):125141, 2022.

\bibitem{silversmith1985zeeman}
A.~J. Silversmith and A.~P. Radlinski.
\newblock {Z}eeman spectroscopy of the {G1} centre in {CaF$_{2}$:Eu$^{3+}$}.
\newblock {\em Journal of Physics C: Solid State Physics}, 18(22):4385, 1985.

\bibitem{Bartholomew}
J.G. Bartholomew.
\newblock {\em Investigation of the scalability of rare-earth-ion quantum
  hardware}.
\newblock PhD thesis, Australian National University, 2014.

\bibitem{liu2005}
Guokui Liu.
\newblock Electronic {Energy} {Level} {Structure}.
\newblock In Guokui Liu and Bernard Jacquier, editors, {\em Spectroscopic
  {Properties} of {Rare} {Earths} in {Optical} {Materials}}. Springer Science
  \& Business Media, 2006.

\bibitem{Reid2006spectroscopic}
M.~F. Reid.
\newblock Transition {Intensities}.
\newblock In G.~Liu and B.~Jacquier, editors, {\em Spectroscopic {Properties}
  of {Rare} {Earths} in {Optical} {Materials}}. Springer Science \& Business
  Media, 2006.

\bibitem{ReRi83c}
M.~F. Reid and F.~S. Richardson.
\newblock Electric dipole intensity parameters for lanthanide {4$f$--4$f$}
  transitions.
\newblock {\em J. Chem. Phys.}, 79:5735--5742, 1983.

\bibitem{ReDaRi83}
M.~F. Reid, J.~J. Dallara, and F.~S. Richardson.
\newblock Comparison of calculated and experimental{4$f$--4$f$} intensity
  parameters for lanthanide complexes with isotropic ligands.
\newblock {\em J. Chem. Phys.}, 79:5743--5751, 1983.

\bibitem{Reid&Richardson}
M.F. Reid and F.S. Richardson.
\newblock Electric dipole intensity parameters for lanthanide {4f $\rightarrow$
  4f} transitions.
\newblock {\em J. Chem. Phys.}, 79(12):5735--5742, 1983.

\bibitem{NN89a}
D.~J. Newman and Betty Ng.
\newblock The superposition model of crystal fields.
\newblock {\em Rep. Prog. Phys.}, 52:699--763, 1989.

\bibitem{NeNg00}
D.~J. Newman and B.~K.~C. Ng, editors.
\newblock {\em Crystal Field Handbook}.
\newblock Cambridge University Press, Cambridge, 2000.

\bibitem{ReNgNe89}
M.~F. Reid, B.~Ng, and D.~J. Newman.
\newblock {\it Ab-initio} calculation of intensity parameters for the system
  {Pr$^{3+}$--Cl$^{-}$}.
\newblock {\em J. Less Common Metals}, 148:219--222, 1989.

\bibitem{Wen2014}
J.~Wen, M.~F. Reid, L.~Ning, J.~Zhang, Y.~Zhang, C.-K. Duan, and M.~Yin.
\newblock Ab-initio calculations of {Judd–Ofelt} intensity parameters for
  transitions between crystal-field levels.
\newblock {\em Journal of Luminescence}, 152:54–57, August 2014.

\bibitem{Dieke:63}
G.~H. Dieke and H.~M. Crosswhite.
\newblock The spectra of the doubly and triply ionized rare earths.
\newblock {\em Appl. Opt.}, 2(7):675--686, 1963.

\bibitem{phillips}
W.L. Phillips~Jr. and J.E. Hanlon.
\newblock Oxygen penetration into single crystals of calcium fluoride.
\newblock {\em J. Am. Cer. Soc.}, 46(9):447--449, 1963.

\bibitem{Carnall}
W.T. Carnall, G.L. Goodman, K.~Rajnak, and R.S. Rana.
\newblock A systematic analysis of the spectra of the lanthanides doped into
  single crystal {LaF$_3$}.
\newblock {\em J. Chem. Phys.}, 90(7):3443--3457, 1989.

\bibitem{ReidCor}
M.F. Reid.
\newblock Correlation crystal field analyses with orthogonal operators.
\newblock {\em J. Chem. Phys.}, 87(5):2875--2884, 1987.

\bibitem{BuJaRiRe94}
G.~W. Burdick, C.~K. Jayasankar, F.~S. Richardson, and M.~F. Reid.
\newblock Energy-level and line-strength analysis of optical transitions
  between stark levels in {Nd$^{3+}$:Y$_3$Al$_5$O$_{12}$}.
\newblock {\em Phys. Rev. B}, 50:16309--16325, 1994.

\bibitem{DaReRi84}
J.~J. Dallara, M.~F. Reid, and F.~S. Richardson.
\newblock Anisotropic ligand polarizability contributions to intensity
  parameters for the trigonal {Eu(ODA)$^{3\hbox{-}}_{3}$} and
  {Eu(DBM)$_{3}$H$_{2}$O} systems.
\newblock {\em J. Phys. Chem.}, 88:3587--3594, 1984.

\end{thebibliography}

\end{document}